# Co-Located Magnetic Levitation Haptic and Graphic Display using Iron Core Coils under Screen

Peter Berkelman, Steven Kang, Sean Trafford, and Muneaki Miyasaka

*Department of Mechanical Engineering, University of Hawaii, Honolulu, United States*

(Email: peterb@hawaii.edu)

**Abstract ---** This paper describes a combined haptic and graphical interactive system in which a grasped handle is levitated and controlled so that its dynamic rigid-body motion and the forces and torques generated upon it match those of a tool in a real-time simulated environment, displayed on a thin screen on top of the levitation coils and underneath the levitated handle. In this augmented reality configuration, the haptic sensations delivered to the hand of the user and the displayed simulation graphics are perceived in the same location, and the graphical display of the tool acts as a virtual extension of the grasped handle into the displayed simulated environment. The novelty of the system is that it combines iron core levitation coils with a low-cost position sensing system and co-located display in a portable system. The high closed-loop control bandwidth and precise position sensing of the system enable interactive simulated environments to be presented with a convincing degree of realism. The interactive environments to be demonstrated will include 3D rigid-body dynamics, surface contacts with stiffness and damping, and surface texture and friction.

## 1 Introduction

We plan to demonstrate a system which provides compelling realistic physical interaction with dynamic simulated environments through the use of an electromagnetic levitation system and a co-located graphical display, placed in between the array of fixed actuation coils and the levitated handle. In this configuration, the motion of a tool in the simulated environment follows the grasped handle while the interaction forces and torques from the simulation are generated onto the handle to be felt by the user. The aim of the control and simulation software is for the virtual tool displayed in the simulation to be perceived as a direct extension of the actual handle grasped by the user, as an augmented reality system. The advantage of magnetic levitation for this co-located interaction is that there are no mechanical parts attached to the levitated handle which would occlude the user's view of the simulated environment on the screen. The novel features of the present implementation are that iron core levitation coils are used, increasing the maximum levitation height and the maximum haptic interaction forces and torques which can be generated on the levitated handle, and that a compact position tracking system is integrated with the levitation coils and graphical display.

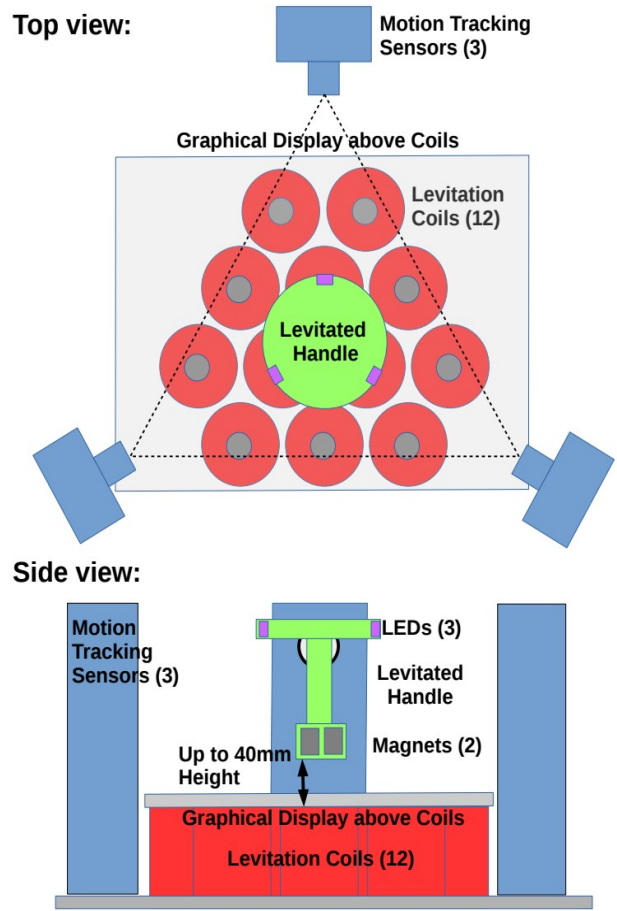

Fig. 1 System Hardware

A schematic view of the system hardware is shown in Fig. 1. Non-contact forces and torques can be generated on the levitated handle in any direction by an array of 12 cylindrical coils with iron cores. Feedback control and motion tracking are used to align and synchronize the motion of the levitated handle with the simulated tool. Position sensing is done using three planar position sensitive diodes on the system base to detect and track three infrared light-emitting diodes (LEDs) fixed to the levitated handle so that the handle's spatial position and orientation can be calculated by triangulation.

## 2 Related Previous Work

### 2.1 Co-Located Haptics and Graphics

Surface-based haptics has been an active topic of research as a means to provide haptic cues associated wth displayed images [1]. Advantages of co-located haptics and graphics in terms of perception, responsiveness, and effectiveness have been surveyed and studied in [2] and [3]. Example co-located systems are described in [4] and [5]. Earlier co-located haptic and graphic systems using magnetic levitation from our lab are shown in Fig. 2. The present system combines co-located haptics and graphics with a compact motion tracking system for position feedback, resulting in a self-contained portable complete system.

### 2.2 Coil Array Magnetic Levitation and Haptics

The design, modeling, and control methods used for long-range magnetic levitation with arrays of cylindrical coils are described in [6] and [7], and sample levitation systems using these methods are shown in Fig. 3. A coil array magnetic levitation system with a novel low-cost, compact non-contact motion tracking system is described in [8] and shown in Fig. 4. The present system uses iron core coils rather than the non-ferrous coil cores used previously, which necessitates more sophisticated actuation and control modeling to account for nonlinear magnetic induction and saturation effects in the iron cores.

## 3 Actuation

Numerical electromagnetic models and matrix transformation methods are used to calculate the coil currents needed to generated the desired forces and torques at each control update, as described in [7]. The novelty of the actuation system to be demonstrated is in the use of cylindrical coils with iron cores rather than the non-ferrous cores used previously. Iron core coils greatly increase the magnetic fields and actuation forces and torques obtained from the coils due to induced magnetization, but complex nonlinear effects in force and torque generation may arise from variations in induced magnetization in the cores from nearby permanent magnets and neighboring coils. These effects must be minimized by careful design principles and by the compensation of the feedback control system.

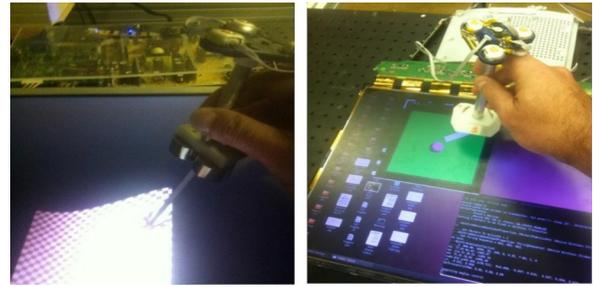

Fig. 2 Co-Located Haptic and Graphic Interaction

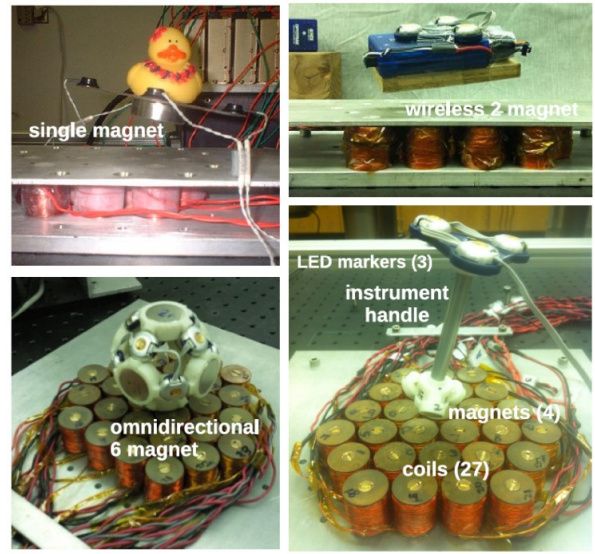

Fig. 3 Magnetic Levitation with Planar Arrays of Cylindrical Coils

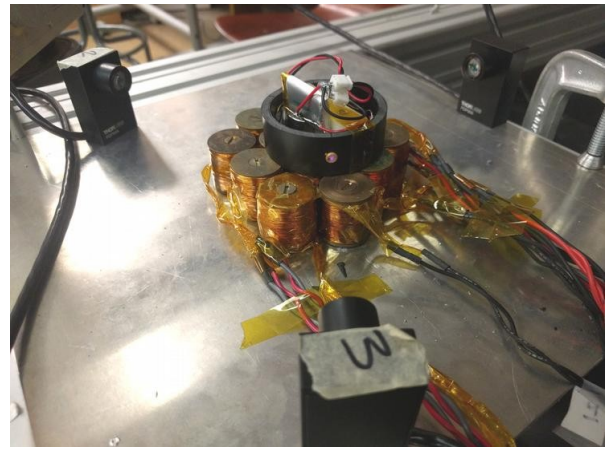

Fig. 4 Coil Array and Position Sensors with Levitated Platform

The use of cylindrical coils and magnets simplifies the actuation models due to radial symmetry. The use of two magnets in the levitated handle enables yaw torques to be generated, because a single levitated cylindrical magnet would spin freely with no means to control its rotation about its cylindrical axis.

The coil and magnet parameters to be used in the demonstration are listed in Table 1. Each coil is driven by a current amplifier (Copley Controls 4121A). Levitation heights up to 40 mm and 3-4 N haptic forces have been measured using these parameters.

Table 1  Coil and Magnet Parameters

| Coils (12): | |
|---|---|
| Height: | 27 mm |
| Outer Diameter: | 25 mm |
| Inner Diameter: | 12.5 mm |
| Windings: | 1000 |
| Coil Separation: | 2 mm |
| Maximum Current: | 4.0 A |
| **Coil Cores (12):** | |
| Height: | 37 mm |
| Diameter: | 8 mm |
| Material: | Alloy Steel |
| **Magnets (2):** | |
| Thickness: | 9.02 mm |
| Diameter: | 19.05 mm |
| Material: | NdFeB |
| Magnetization: | N52 |

### 4  POSITION SENSING

Real-time measurement of the handle spatial position and orientation is necessary for accurate force and torque generation and for feedback control of the dynamic interaction between the levitated handle and the simulated environment. Infrared LED position markers are attached to the top of the levitated handle, facing in different directions so that each LED is visible to only one photodiode sensor (Thorlabs PDP90A). A lens in front of each position sensing photodiode images the light from the corresponding LED onto the plane of the diode to indicate the horizontal and vertical angles from each sensor to its visible LED. The 3D position and spatial orientation of the levitated handle can then be calculated from the six sensor angles by geometric triangulation and an incremental error minimization procedure. Use of these low-cost, compact position sensors replaces much larger and more costly optical motion tracking systems.

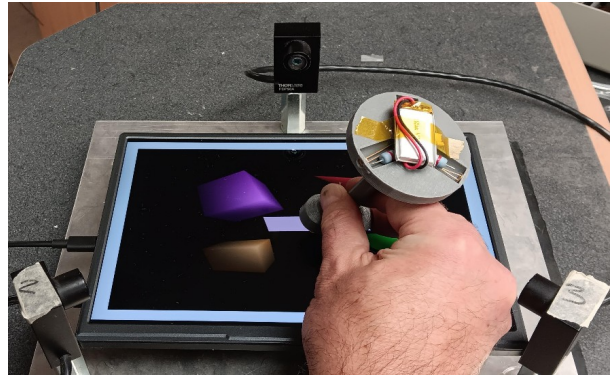

Fig. 5  Graphical Display, Position Sensors, and User Handle with Magnets and LEDs

A battery to drive the infrared LEDs is also attached to the levitated body to enable completely wireless operation. The 3D-printed levitated handle for co-located interaction is shown in Fig. 5 with the position sensors and graphical display.

### 5  CONTROL AND INTERACTIVE SIMULATION

Forces and torques to be generated on the grasped handle, and its sensed position and orientation are calculated by the control and interactive simulation software at an update rate up to 2 kHz. Trajectory following motion control of the levitated handle can be implemented by a control law which includes a constant feedforward vertical force term to cancel the force of gravity, and proportional and derivative error feedback force and torque terms to servo to desired trajectory setpoints in each of the six degrees of freedom in rigid-body translation and rotation. Motion control bandwidths of at least 200 Hz are possible with motion ranges of approximately 30 mm vertical, 80 mm horizontal, ±45 degrees of yaw, and ±30 degrees of tilt in roll and pitch. Digital control software is implemented on a Linux PC with PCI analog conversion boards (Advantech PCI-1716 and PCI-1724U).

For human interaction, user feedback forces and torques are calculated according to the dynamic simulation instead of by feedback motion control laws. At each control update, the interpenetration between the simulated tool and other rigid-body modeled objects in the simulation is calculated and haptic feedback forces and torques are calculated according to proportional and derivative feedback gains applied to the magnitude, position, and direction of the calculated interpenetration volumes. Friction and surface texture models applied to the haptic feedback increase the realism of the interaction. The motion of other moving objects in the simulation is also updated according to discrete time Newtonian dynamics and the virtual forces and torques acting upon them. The high motion and force control bandwidths from noncontact electromagnetic actuation instead of a motorized linkage enable interactive haptics at vibrotactile frequencies, so that realistic and detailed surface textures can be displayed without the use of additional vibrotactile actuators.

The graphical display of the 3D simulated environment is updated at the frame rate of the display. The nominal 8 mm thickness of the display screen, subtracted from the maximum 40 mm levitation height, provides a motion range for 3D interactive motion up to 32 mm above the screen surface.

## 6 Conclusion

The co-located magnetic levitation haptic and graphical interface system will be demonstrated with multiple interactive 3D dynamic physical simulations including moving objects, solid contacts, mass, stiffness, and damping, surface texture, and friction. The actuation, display, control, and interaction methods used can be scaled to larger areas without difficulty, so that haptic interactive full-body medical simulations are feasible as future works. We also plan to use the present system as a testbed to study the effects of realistic co-located haptic and graphic interaction on human perception and task performance in virtual environments.